# Lattice Properties of $MgB_2$ versus Temperature and Pressure

J. D. Jorgensen, D. G. Hinks, and S. Short,
Materials Science Division, Argonne National Laboratory, Argonne, IL 60439

ABSTRACT

We have determined the structural properties of the superconducting compound $MgB_2$ as a function of temperature from 11 K to 297 K and as a function of hydrostatic pressure up to 0.62 GPa using neutron powder diffraction. This compound, when compared to other diborides with the same structure, is characterized by unusually large anisotropies of both the thermal expansion and compressibility, with the *c*-axis responses being substantially larger in both cases. We speculate that the comparatively weaker metal-boron bonding in $MgB_2$, manifest by these lattice responses, is important for establishing the structural features that give rise to high $T_c$ superconductivity in this structure type.

The remarkable discovery of superconductivity at 39 K in $MgB_2$ [1] illustrates the critical role of electronic and crystal structure properties in achieving superconductivity in a given structure type. $MgB_2$ possesses the simple hexagonal $AlB_2$-type structure (C32 structure) [2,3], which is perhaps the most common structure type among the borides.[4] This structure type has previously been investigated extensively for superconductivity. In 1970, Cooper et al. [5] reported a search for superconductivity in this structure type in the series of compounds $YB_2$-$ZrB_2$-$NbB_2$-$MoB_2$. They worked under the assumption that the compounds could be non stoichiometric and investigated boron-poor and boron-rich materials as well as solid solutions. They were able to achieve superconducting transition temperatures, $T_c$, of 3.87 K in a "boron-rich" $NbB_2$ compound and above 11 K in $Zr_{0.13}Mo_{0.87}B_2$. Their experimental approach was that variations in stoichiometry or chemical composition could be used to produce superconductivity by adjusting the formal electrons-per-atom (e/a) count to what they considered an optimal value of 3.8. In light of what is now being learned about superconductivity in $MgB_2$, it is interesting that the highest $T_c$s seen by Cooper et al. using this approach also corresponded to maximizing the length of the *c* axis. In 1979, Leyarovski and Leyarovski [6] searched again for superconductivity in compounds with the $AlB_2$-type structure. They investigated $MeB_2$ (Me=Ti, Zr, Hf, V, Nb, Ta, Cr, and Mo) and found superconductivity only in $NbB_2$, which displayed a $T_c$ of 0.62 K for their sample.

Following the discovery of superconductivity in $MgB_2$ [1], Slusky et al. [7] were the first to report studies of how the behavior responds to incremental changes in chemical composition. They investigated the substitution of Al on the Mg site, i.e., $Mg_{1-x}Al_xB_2$, and observed that $T_c$ decreases smoothly with increasing x for $0 \leq x \leq 0.1$, accompanied by a slight decrease of the *c* axis. At $x \approx 0.1$, there is an abrupt transition to a non-superconducting isostructural compound which has a *c* axis shortened by about 0.1 Å. They concluded that the compound $MgB_2$ is near a structural instability, at slightly higher electron concentration, that can destroy superconductivity. Clearly, the loss of superconductivity associated with decreasing the *c* axis length with no change in cell symmetry and only a small change in the formal electron count suggests that there is something special about the structural parameters of $MgB_2$ that leads to superconductivity in this compound. The importance of the *c* axis length is reminiscent of the earlier work of Cooper et al.[5]

An and Pickett [8] calculated the effects of various phonon modes on the electronic structure of $MgB_2$. They concluded that the superconductivity results almost exclusively from B $\sigma$ bands that contribute strongly to the Fermi level density of states because of the two-dimensional nature of the compound. Their calculations showed that B in-plane phonons, the $E_{2g}$ modes at a calculated [9] energy of 58 meV, are strongly coupled to this band. They explained the instability observed by Slusky et al.[7] in terms of the disappearance of this band upon doping. These results suggest that both the appropriate structure, with a sufficient two-dimensional character which gives rise to the presence of significant density of states in the B σ band, and the correct electron count, to place this band near the Fermi energy, are required to achieve superconductivity in this structure type.

In this paper, we report the structural parameters of $MgB_2$ as a function of temperature from 11 K to 297 K and as a function of hydrostatic pressure, at room temperature, to 0.62 GPa and compare the behavior with that of other compounds with the same structure type. The thermal expansion can be nicely modeled with a simple Einstein function using a single phonon energy around 500 K (43 meV). This is typical of materials with a stiff elastic response. Structural measurements vs. pressure are of interest because the application of pressure allows the structure to be changed continuously, modifying the electronic structure, phonon frequencies, and electron-phonon coupling, without changes in chemical composition or formal electron count. Very recent measurements show that the superconducting transition temperature, $T_c$, decreases with the application of pressure [10,11] as is generally predicted for an electron-phonon superconductor resulting from the pressure-induced increases of phonon frequencies.[12] A knowledge of the changes in structure vs. pressure, which we observe to be markedly anisotropic, is needed to quantitatively interpret this behavior.

Because the isotope $^{10}B$ has a large neutron absorption cross section, a 1.6 g sample for this study was made using isotopically-enriched $^{11}B$ (Eagle Picher, 98.46 atomic % enrichment). A mixture of $^{11}B$ powder (less than 200 mesh particle size) and chunks of Mg metal was reacted in a capped BN crucible at 800 C under an argon atmosphere of 50 bar for 1.5 hours. The resulting sample displayed a sharp superconducting transition (0.4 K wide) with an onset at 39 K. Both x-ray and neutron diffraction data showed the sample to be single phase with the $AlB_2$-type structure.

Neutron powder diffraction measurements were made on the Special Environment Powder Diffractometer at the Intense Pulsed Neutron Source, Argonne National Laboratory.[13] For the low-temperature measurements, the sample was contained in a sealed thin-walled vanadium can along with helium exchange gas and cooled using a Displex refrigerator. The measurements vs. pressure were made in a helium gas pressure cell [14], at room temperature. Typical data collection times were 1 hour at each temperature or pressure, except for a small number of longer runs (up to 4 hours) to look closely for any possible superlattice peaks. The data were analyzed by the Rietveld technique using the GSAS code.[15] In initial refinements, the Mg/$^{11}B$ ratio was refined. There was no indication of non stoichiometry within a refinement precision of about 0.5%. Fig. 1 shows the raw data and refined diffraction pattern at 34 K for a 1-hour data collection. The sample is single phase and the diffraction pattern is nicely fit with peak widths near the instrumental resolution. This is true at all temperatures and pressures. There is no evidence for any structural transitions. Refined structural parameters at 297 K and 37 K are listed in Table 1.

The simple hexagonal $AlB_2$-type structure (space group P6/mmm, No. 191) [4] is shown in Fig. 2. The structure contains graphite-like boron layers which are separated by hexagonal close-packed layers of metals. The center of a hexagonal boron ring lies both directly above and below each metal.

The lattice parameters and cell volume vs. temperature are shown in Fig. 3. The thermal expansion can be nicely modeled with an Einstein equation using a single phonon energy:

$$\ln\left(\frac{a}{a_0}\right) = \frac{A\theta}{e^{(\theta/T)} - 1}, \qquad (1)$$

where a is the lattice parameter (*a* or *c*) or cell volume (*V*) and $a_0$ is its value at T=0, θ is the phonon energy, T is the temperature, and A is a scaling coefficient. The data of Fig. 3 have been fit with this equation to determine the values of $a_0$, A, and θ. Independent fits to *a*, *c*, and *V* give the same phonon energy, θ, within the standard deviations: 517(20) K, 494(12) K, and 508(13) K for the fits to *a*, *c*, and *V*, respectively.

The thermal expansion along the *c* axis is about twice that along the *a* axis. Linear thermal expansions near room temperature can be approximated from the data for 200 K≤T≤300 K, giving $\alpha_a \approx 5.4\text{x}10^{-6}$ $K^{-1}$ and $\alpha_c \approx 11.4\text{x}10^{-6}$ $K^{-1}$ [where α is defined as $(\Delta l/\Delta T)/l_0$]. The thermal expansions (above room temperature) of a number of other diboride compounds with the same structure have been published.[14] The thermal expansion of $MgB_2$ lies generally within the range observed for these other diborides, although only $VB_2$ (for which $\alpha_c=14\text{x}10^{-6}$ $K^{-1}$ at room temperature) has a larger *c*-axis thermal expansion. Larger thermal expansion along the *c* axis than along the *a* axis is not unusual for the $AlB_2$-type diborides, but most do not display such a large anisotropy as does $MgB_2$. This anisotropy results from the difference in bond strengths. The B-B bonds in the basal plane are much stronger than the Mg-B bonds that connect layers of Mg and B atoms.

Thermodynamics requires a small positive response of the lattice parameters at $T_c$ that is often beyond the limits of sensitivity of conventional diffraction measurements.[16] The details of our lattice parameter measurements at low temperature are shown in the insets of Fig. 3. Small positive effects may be seen near $T_c$ for the cell volume and the *a*-axis lattice parameter, but these are admittedly within the experimental uncertainties. A more sensitive technique, such as higher resolution diffraction or dilatometry, will be required to properly investigate this behavior. However, in light of the conclusions of An and Pickett [8] that the in-plane B $\sigma$ bands are responsible for the superconductivity, it is intriguing that we may observe a tiny structural response in the basal plane, but nothing perpendicular to the basal plane. This could be a manifestation of the change in B-B bonding as the bonding electrons condense at $T_c$.

The refined Debye-Waller factors in the basal plane and perpendicular to the plane for Mg and B vs. temperature are shown in Fig. 3. The larger vibrational amplitudes along the *c* axis than along the *a* axis for the B atoms are another manifestion of the weaker Mg-B bonding. The mean squared displacements decrease as expected, but have significantly non-zero values at low temperature, especially for $U_{33}$(B). This could result from the inability of the Debye-Waller factor model of the Rietveld code (which assumes harmonic thermal vibrations) to fit anharmonic behavior, or a small static displacement of the B atoms that persists to low temperatures. Such displacements, if ordered, would give rise to supercells of the basic $AlB_2$-type structure. No evidence for such supercells is visible in our data.

The variation of the *a* and *c* lattice parameters vs. pressure is shown in Fig. 4. Over the pressure range of this study, the changes are linear and can be expressed as

$$a = a_0 - 0.00187P, \text{ and}$$

$$c = c_0 - 0.00307P,$$

where $a_0$ and $c_0$ are the zero-pressure lattice parameters and P is the pressure in GPa.

Compression along the *c* axis is 64% larger than along the *a* axis, consistent with the comparatively weaker (Mg-B) bonds that determine the *c* axis length. A similar anisotropy, but not as large, has been reported in the refractory diboride $TiB_2$ [17], which is of considerable technological interest because of its high elastic moduli, high hardness, and high electric conductivity. By comparison, the compression anisotropy in the layered cuprate is about a factor of two.[14] The markedly anisotropic compression behavior of $MgB_2$ will lead to different pressure effects on different phonon modes and is also more likely to lead to changes in the electronic structure at the Fermi energy than when compression is isotropic because B-B and Mg-B distances change at different rates. The markedly anisotropic compression also may pose an experimental difficulty in characterizing pressure effects in this material, because the use of non-hydrostatic pressure transmitting media is more likely to produce shear and lead to incorrect results.

Sluskly et al. [4] showed that the substitution of Al on the Mg site decreases the *c* axis at a rate approximately twice that observed for the *a* axis. The similarity of the structural effect to that produced by the application of pressure is intriguing, even though the substitution of Al is also changing the electron count. $T_c$ decreases with Al substitution. At approximately 10% Al substitution, they observe an abrupt isostructural transition with an additional reduction of the *c* axis and superconductivity is lost. They attribute this to an instability that destroys superconductivity. Very similar structural effects, without the change in electron count, are produced by the application of pressure. Extrapolating our results, a compression along the *c* axis equivalent to the substitution of 10% Al would require approximately 2.6 GPa hydrostatic pressure. Measurements of $T_c$ vs. pressure have not yet been extended to this pressure range. Measurements to higher pressure could provide a test of the conclusion of An and Pickett [8] that changes in the electron count resulting in the disappearance of the $\sigma$ band are responsible for the sudden *c*-axis collapse. It may be possible to use pressure to significantly alter the character of the $\sigma$ band, as the structure becomes less two-dimensional, without eliminating it.

The structural data vs. temperature and pressure reported here, when considered in the context of previous efforts to achieve and characterize superconductivity in the $AlB_2$-type structure and what is currently being learned about $MgB_2$, allow us to speculate about what structural features are important for achieving superconductivity in this structure type. Somewhat weak metal-boron bonding, which is manifest by a comparatively long *c* axis, larger thermal expansion and compression along the *c* axis, and larger vibrational amplitudes along the *c* axis, is a characteristic feature of $MgB_2$ and leads to its somewhat two-dimensional-like electronic structure that is thought to be of critical importance.[8] This relatively weak bonding is to be expected when a divalent metal is used in the hexagonal metal diboride structure, in contrast to the compounds based on transition metals that have been most extensively studied. The early work on hexagonal diborides by Cooper et al. [5] and the very recent work of Slusky et al. on $Mg_{1-x}Al_xB_2$ [7] are consistent with this hypothesis. Following the logic that has typically been applied to raise $T_c$ in a given structure type, researchers will undoubtedly focus attention on changing the electronic structure such that the Fermi energy is positioned at a peak in the electronic density of states. As the search for higher $T_c$s in compounds with the $AlB_2$-type structure proceeds, it will be important to realize that chemical/structural changes that modify the metal-boron bonding strength could be unusually important in these compounds.

ACKNOWLEDGEMENT

This work was supported by the U. S. Department of Energy, Office of Science, contract No. W-31-109-ENG-38.

Table 1. Refined structural parameters for $MgB_2$ at 297 and 37 K based on Rietveld refinements using neutron powder diffraction data. Space group P6/mmm, No. 191, with Mg at (0, 0, 0) and B at (1/3, 2/3, 1/2). Numbers is parenthesis are statistical standard deviations of the last significant digit.

| | 297 K | 37 K |
|---|---|---|
| a (Å) | 3.08489(3) | 3.08230(2) |
| c (Å) | 3.52107(5) | 3.51461(5) |
| V (Å$^3$) | 29.019(1) | 28.917(1) |
| $U_{11}$(Mg) (Å$^2$) | 0.00545(26) | 0.00347(18) |
| $U_{33}$(Mg) (Å$^2$) | 0.00559(42) | 0.00328(29) |
| $U_{12}$(Mg) (Å$^2$) | 0.00272(13) | 0.00173(9) |
| $U_{11}$(B) (Å$^2$) | 0.00454(15) | 0.00333(11) |
| $U_{33}$(B) (Å$^2$) | 0.00648(25) | 0.00455(18) |
| $U_{12}$(B) (Å$^2$) | 0.00227(8) | 0.00166(6) |

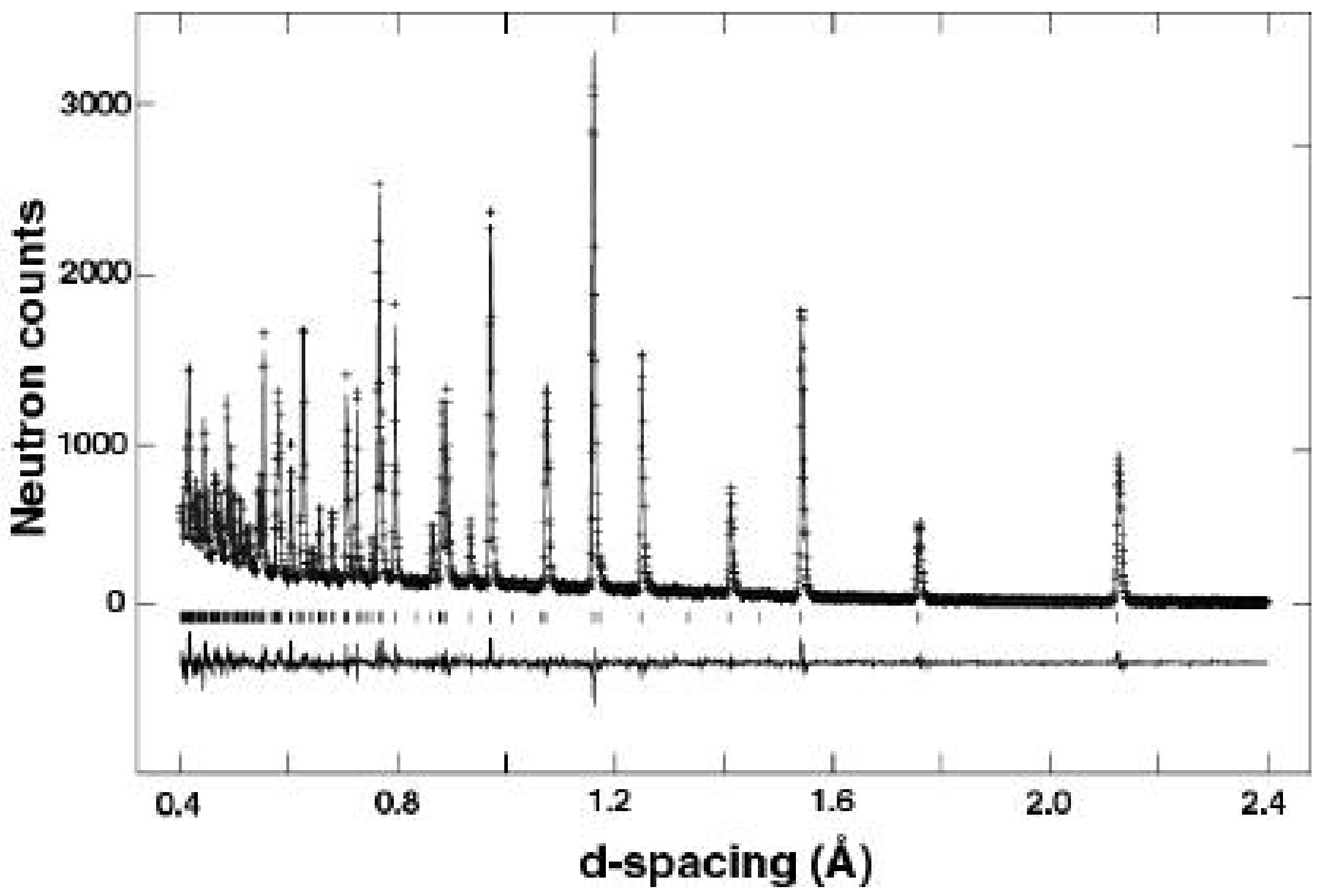

Fig. 1. Observed neutron powder diffraction data and best-fit Rietveld refinement profile for $MgB_2$ at 34 K. Data collection time was 1 hour. Crosses (+) are the raw data. The solid line is the calculated profile. Tick marks indicate the positions of all allowed reflections. A difference curve (observed minus calculated) is plotted at the bottom.

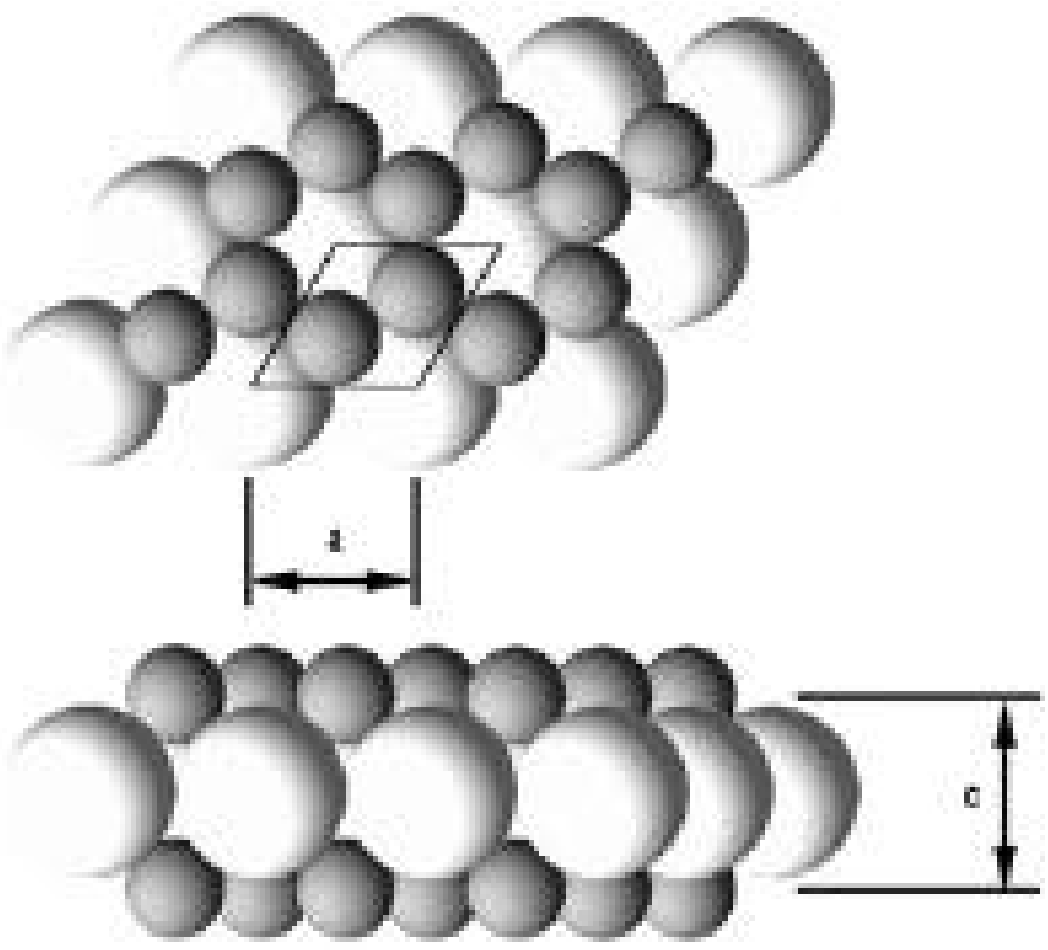

Fig. 2. Crystal structure of $MgB_2$ [$AlB_2$-type structure; hexagonal space group P6/mmm, No. 191, with Mg at (0, 0, 0) and B at (1/3, 2/3, 1/2)] viewed along the *c* axis (top) and perpendicular to an *a* axis (bottom). Small spheres are B atoms; larger spheres are Mg atoms.

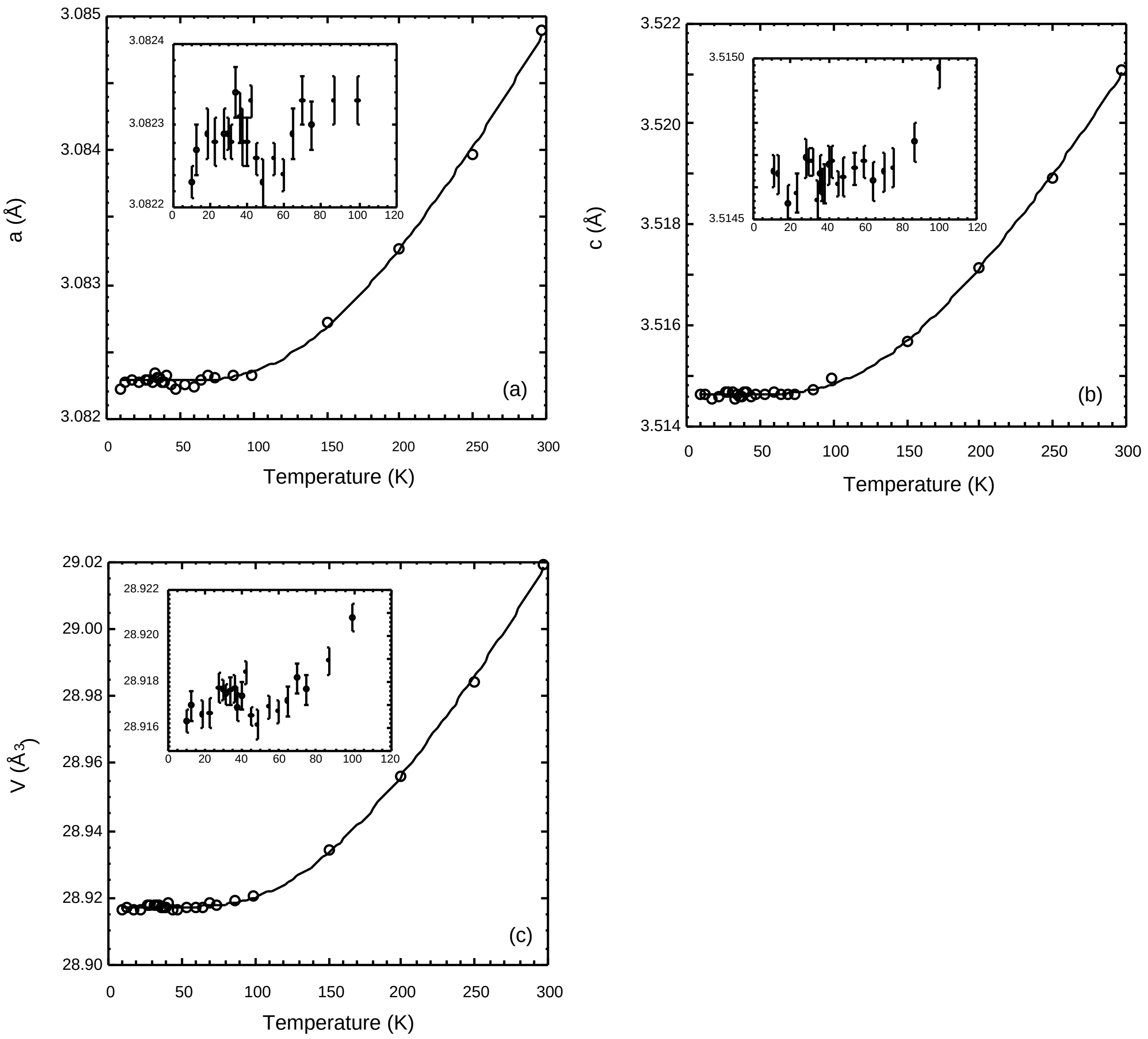

Fig. 3. *a* and *c* lattice parameters and unit cell volume, V, of $MgB_2$ vs. temperature based on neutron powder diffraction measurements. The solid lines are least squares fits using a simple Einstein model with a single phonon energy [equation (1) in the text] and yield phonon energies of 517(20) K, 494(12) K, and 508(13) K for the fits to *a*, *c*, and V, respectively. The insets show the low-temperature data in more detail. Where not shown, standard deviations are smaller than the symbols.

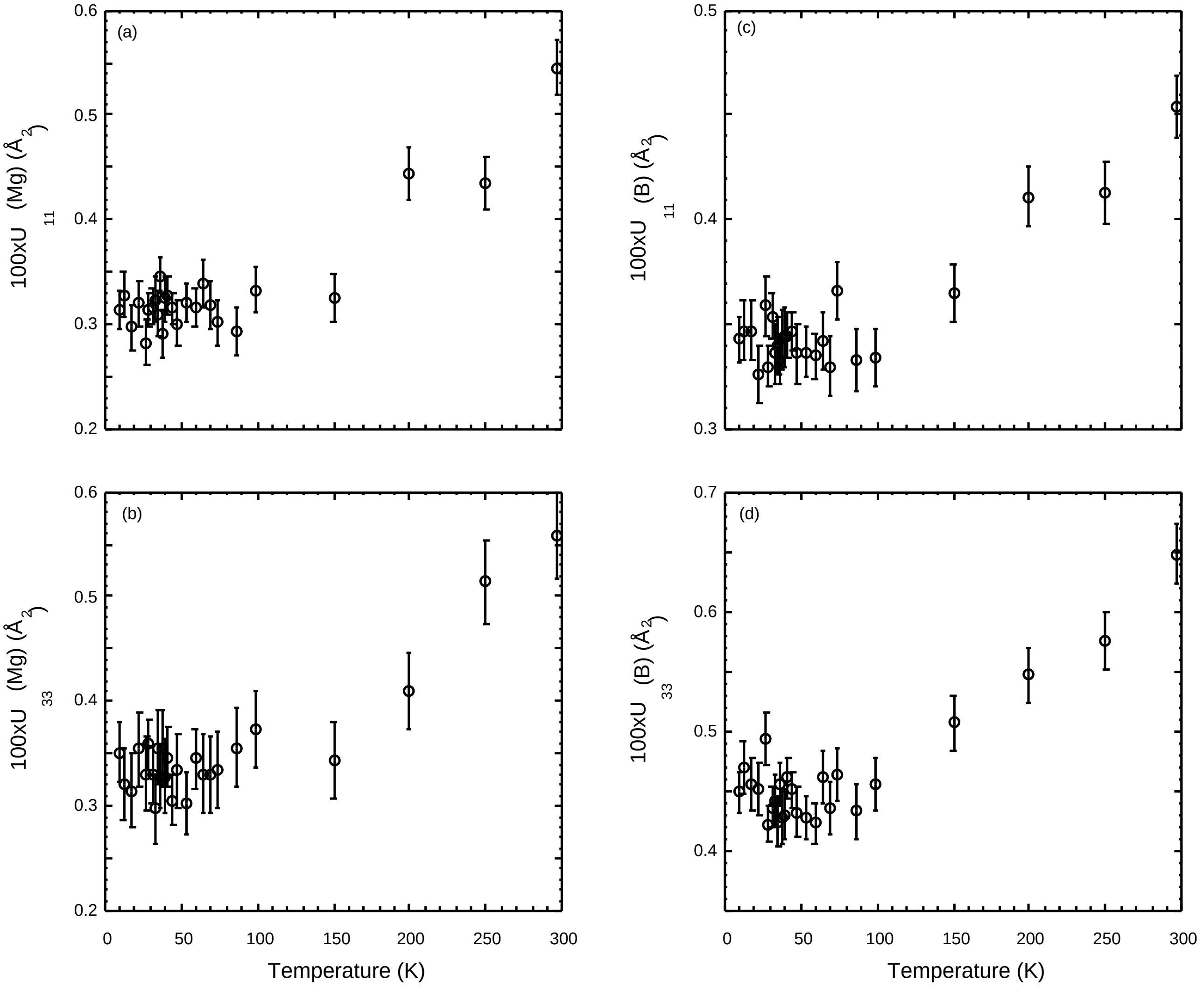

Fig. 4. Diagonal components, $U_{11}$ and $U_{33}$, of the anisotropic Debye-Waller tensors for Mg (a and b) and B (c and d) atoms in $MgB_2$ vs. temperature determined by Rietveld refinement using neutron powder diffraction data.

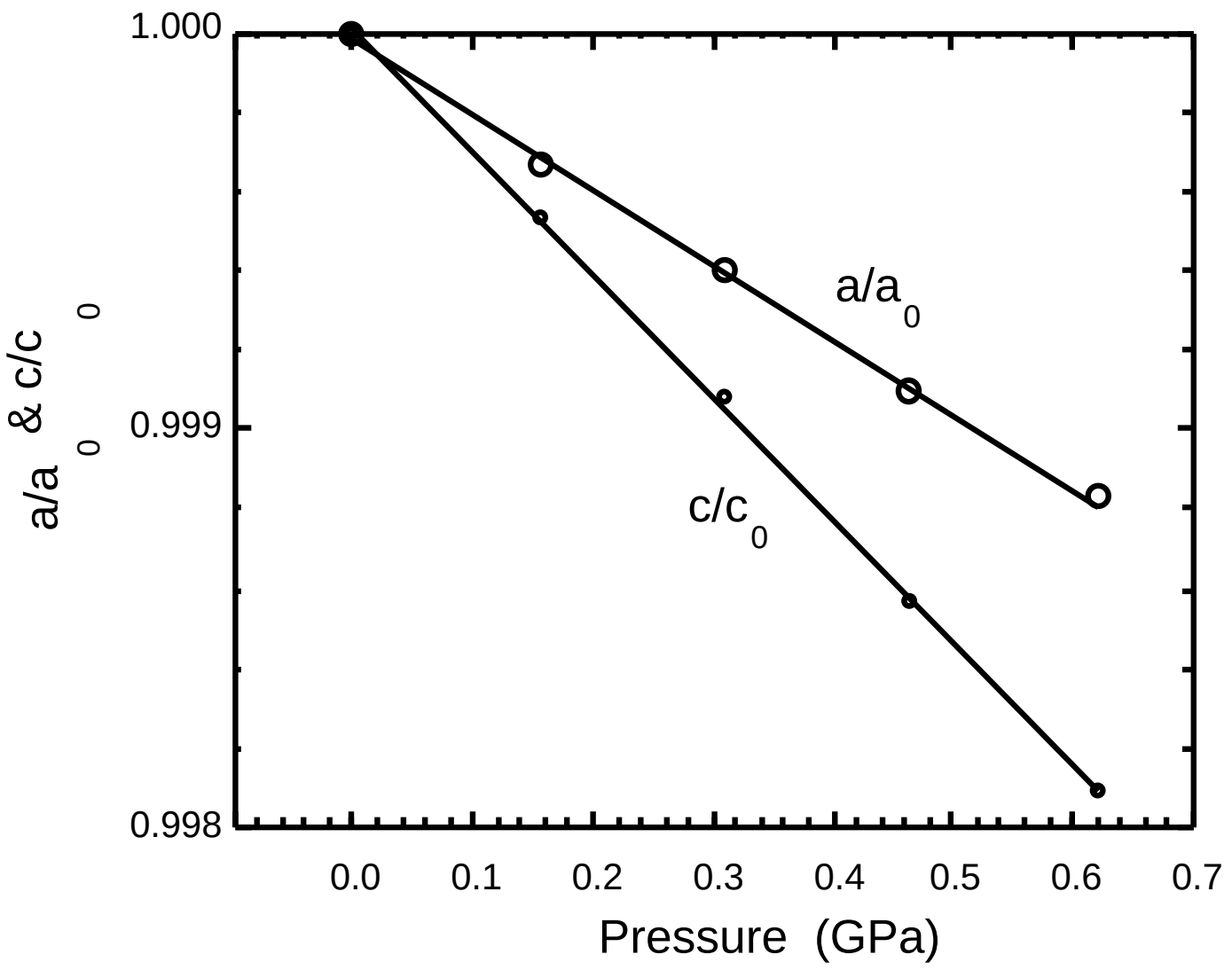

Fig. 5. Normalized *a* and *c* lattice parameters vs. pressure at room temperature for $MgB_2$ based on neutron diffraction measurements at five pressures using helium as the pressure transmitting medium. Standard deviations of the individual points are smaller than the symbols. The straight lines are linear least-squares fits to the data